\documentclass[journal]{IEEEtran}
\usepackage{graphicx} 
\usepackage{amsmath}
\usepackage{amsfonts}
\usepackage{amssymb}
\usepackage{listings}
\usepackage{algorithmic}
\usepackage[english]{babel}
\usepackage{algorithm}
\usepackage[small]{caption}
\usepackage[font=footnotesize]{subfig}
\usepackage{color}
\usepackage{multicol}
\usepackage{multirow}
\usepackage{tabularx,colortbl}
\usepackage{setspace}
\usepackage[]{psfrag}
\usepackage{psfrag}
\usepackage{url}
\usepackage{relsize}
\begin{document}
\title{A novel mechanical analogy based battery model for SoC estimation using a multi-cell EKF}
\author{Maurizio~Paschero,
        Gian~Luca~Storti,
        Antonello~Rizzi,
        Fabio Massimo Frattale Mascioli
        and~Giorgio~Rizzoni,~\IEEEmembership{IEEE Fellow}
\thanks{M. Paschero, A. Rizzi and F.M. Frattale Mascioli are with the Information, Electronic and Telecommunication Department (DIET) - POMOS Labs of the University of Rome `Sapienza', Italy.}
\thanks{G.L. Storti and G. Rizzoni are with the Center for Automotive Research (CAR), Ohio State University, Columbus, OH, 43211 USA e-mail: storti.3@osu.edu.}
\thanks{Manuscript received Dec 15, 2015; revised June XX, 2015.}
}
\maketitle
\begin{abstract}
The future evolution of technological systems dedicated to improve energy efficiency will strongly depend on effective and reliable Energy Storage Systems, as key components for Smart Grids, microgrids and electric mobility. 
Besides possible improvements in chemical materials and cells design, the Battery Management System is the most important electronic device that improves the reliability of a battery pack. In fact, a precise State of Charge (SoC) estimation allows the energy flows controller to exploit better the full capacity of each cell. In this paper, we propose an alternative definition for the SoC, explaining the rationales by a mechanical analogy. We introduce a novel cell model, conceived as a series of three electric dipoles, together with a procedure for parameters estimation relying only on voltage measures and a given current profile. The three dipoles represent the quasi-stationary, the dynamics and the istantaneous components of voltage measures. An Extended Kalman Filer (EKF) is adopted as a nonlinear state estimator. Moreover, we propose a multi-cell EKF system based on a round-robin approach to allow the same processing block to keep track of many cells at the same time. Performance tests with a prototype battery pack composed by 18 A123 cells connected in series show encouraging results.
\end{abstract}
\section{Introduction}
\label{sec:intro}
\IEEEPARstart{M}{odern} engineering is already facing the fundamental challenge of improving energetic, environmental and social sustainability in the way energy is produced, distributed, delivered and even consumed by the final users. Many systems and technical components are on the verge of a disruptive transformation, driven by a complex multidisciplinary co-evolution process, where many vital subsystems are closely related one another. This revolution must be faced and performed taking into account a systemic point of view, able to drive technical advancements as sequences of systematic and coherent transformations. Urban areas planning and development, buildings design, electric generation and distribution systems, advanced telecommunication systems, intelligent multimodal transportation systems, cloud computing systems and intelligent processing systems  (just to cite some instances) will benefit each other of advances and technical improvements. Sustainability is the key term to drive and define precise definitions for any objective function acting as the fitness of each technological subsystem in this evolving scenario. 
From this point of view, just to cite an example, the revolution in transportation systems due to the introduction of plug-in electric vehicles (EVs) will contribute in driving the design of next generation of Smart Grids \cite{Yu_2011,deSantis_2015,Storti_2015}, since electric mobility will be a huge additional load for both energy generation and distribution systems. At the same time, the massive introduction of EVs will yield a true reduction of $\mathrm{CO_2}$ emissions only if supported by a model of distributed energy generation from renewable sources. In turn, the stochastic nature of some promising renewable sources (photovoltaics plants and wind generators, for example) demands the presence of stationary energy storage systems (ESSs) to exploit fully the available energy, allowing the spread of Micro Grids. A Micro Grid can in fact be defined as a sub-network characterized by the presence of autonomous (often renewable) energy sources buffered by some type of ESS and locally controlled in order to achieve smart energy flows management \cite{deSantis_2013}. In this scenario, the Smart Grid will evolve into a System of Systems (SoS), where most of the loads and energy sources will be localized into Micro Grids, organized as a hierarchical territorial granulation, and acting as cooperative/competitive agents in a complex energy trading network. The key components for the full deployment of this future technological setting are ESSs. Nowadays the most promising technology is represented by Li-Ion (Lithium Ion) and Li-Po (Lithium Polymer) battery systems, controlled by suited BMSs (Battery Management Systems).
The BMS represents the key component of every chemical ESS, since this electronic device is conceived and designed to manage, protect, monitor, balance and estimate the State of Charge (SoC) of rechargeable batteries. An accurate SoC determination in Li-Ion batteries is the most important aspect for maximizing the battery pack usage and to evaluate and monitor correctly its State of Health (SoH). Determining the exact amount of energy available in a battery is an extremely difficult endeavor, due to the lack of deep knowledge of the electro-chemical behavior of a cell. The only available option is to perform an estimation of the SoC based on an external characterization of the cell, i.e. based on current and voltage measurements. In the literature there exits several techniques to perform this estimation: open circuit voltage (OCV) calculation, coulomb counting, and/or more sophisticated techniques that employ state estimators such as the Extended Kalman Filter (EKF). In the paper, we propose a new battery model and a procedure to characterize a cell, by estimating model’s parameters, finalized to enhance SoC estimation. We explain the procedure relying on a mechanical (hydraulics) analogy. Once performed the parameters estimation, by a suited data driven external characterization, an EKF is employed to lock and track the status of the system. Moreover, we propose a round-robin approach to keep track of the states of multiple cells, by relying on a single implementation of the EKF block. The system is designed to fully exploit all the available computational power of the embedded system running the SoC procedure, and it is based on the fact that in practical applications currents are band-limited signals, characterized by very low maximum frequencies, and thus allowing low sampling rates. 
This paper is organized as follows. In Sect.~\ref{sec::soc} we describe a novel way to define the SoC of a rechargeable cell, while in Sect.~\ref{sec::model} we explain the proposed cell model, defined as a series of three dipoles, each one defined to model contributions affecting the cell behavior at different time scales. Sect.~\ref{sec::parameter} depicts the parameters identification procedure, conceived to isolate and compute the quasi-static, dynamical and istantaneou contributions, by feeding the cell with a suited current profile. The Muli-cell EKF is presented in Sect.~\ref{sec::multi_cell_ekf}. Test setups and results for both the model and the Multi-cell EKF are reported in Sect.~\ref{sec::results}. Finally, our conclusions are drown in Sect.~\ref{sec::conclusion}.

\section{Consderations on the state of charge}
\label{sec::soc}
\subsection{Introduction}
The SoC is a time dependent quantity representing the percentage of the total storable charge still drainable from a cell at a given instant of time \cite{Zou_2014}. Assuming for convenience an unitary efficiency, it is usually defined as
\begin{equation}
SoC(t)=SoC(t_0)+\frac{1}{C_n}\int_{t_0}^{t}I_{in}(\tau)d\tau
\label{eq::oldsoc}
\end{equation}
where $I_{in}$ is the input current and $C_n$ is the nominal capacity of the cell, \textit{i.e.} the  amount of charge drainable, in an hour, from a fully charged cell at a given current rate. A cell is said to be fully charged if a given maximum reference voltage is permanently measured between its terminals in open circuit condition.
Even though \eqref{eq::oldsoc} is quite clear from a theoretical point of view, it is difficult to apply it in actual practice. In fact there are at least three problems with it:
\begin{enumerate}
\item we do not know the SoC value $SoC(t_0)$ at time $t_0$
\item we do not know exactly how the nominal capacity $C_n$, which could be different cell to cell, is related to the maximum and the minimum voltage of the cell.
\item it is very hard to measure the input current $I_{in}$ accurately 
\end{enumerate}
It should be noted that most of the problems listed above are due to the insufficient semantic correlation of \eqref{eq::oldsoc} with the physical parameters of the cell. In fact the only cell parameter involved in \eqref{eq::oldsoc} is $C_n$ and all the differences among cells are flatten in the initial condition $SoC(t_0)$.
In order to overpass these limits, the SoC definition should be tailored cell by cell.
An alternative SoC definition, capable to solve the problems related to \eqref{eq::oldsoc} will be introduced based on a mechanical analogy.
\subsection{A mechanical analogy}
\label{sec::analogy}
The problem of estimating the amount of charge stored in a cell based on voltage and current measurements is quite similar to the problem of estimating the volume of the water stored in a reservoir based on pressure and flow rate measurements.
\begin{figure}[!htbp]
	\centering
		\includegraphics[scale=0.8]{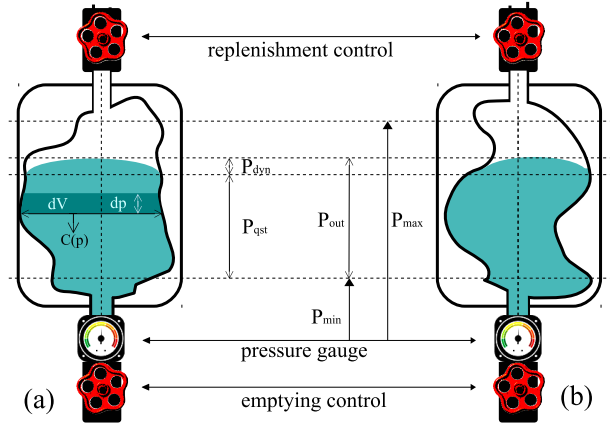}
	\caption{Mechanical analogy of a cell.}
	\label{fig::reservoir}
\end{figure}
As well known, the pressure at a given point is proportional to the height of the column of fluid between this point and the water surface; for this reason from now on we will use the term pressure instead of height. 

It should be noted that different amounts of water stored in two reservoirs having different internal shape can provide the same pressure at the measurement point, as shown in \figurename~\ref{fig::reservoir} parts (a) and (b). In other words, when the internal shape of the reservoir is unknown, it is not possible to determine the amount of water contained in the tank based on pressure measures only.

In fact, the infinitesimal increment in the water volume $dV$ and the consequent pressure variation $dp$ are related one another through the reservoir cross section area $C(p)$ by the relation 
\begin{equation}
C(p)=\frac{dV}{dp}
\label{eq::cp}
\end{equation}
Unfortunately, we don't know the reservoir internal shape, \textit{i.e.} the variation of the internal cross section area $C(p)$ with the pressure measured at the gauge (see \figurename~\ref{fig::reservoir}),
but we can try to estimate it experimentally.
\subsection{Internal shape estimation procedure}
In order to develop a procedure capable to estimate the reservoir internal shape, first of all we need to clarify which actions we can perform on the reservoir. Referring to \figurename~\ref{fig::reservoir}, there are only three actions we can carry out
\begin{itemize}
\item open or close the replenishment control
\item open or close the emptying control
\item read and acquire the pressure $P_{out}$ at the gauge
\end{itemize}
Moreover, we should realize that when we act on the replenishment and the emptying control, the water movement produces the formation of waves on the stationary water level.  
Consequently, the pressure $P_{out}$ read at the gauge will be the summation of a quasi-statical contribution $P_{qst}$ and a dynamical contribution $P_{dyn}$.

\begin{figure}[!htbp]
	\centering
		\includegraphics[scale=0.8]{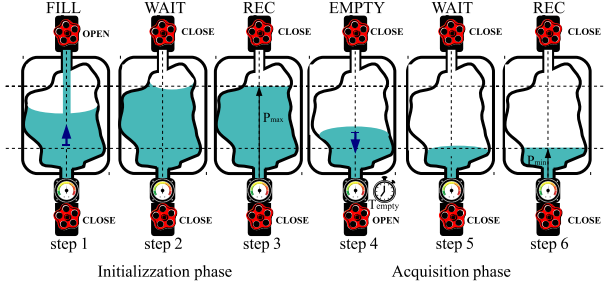}
	\caption{Internal shape extimation procedure.}
	\label{fig::procedure}
\end{figure}

The proposed procedure is composed of two macro phases: an initialization phase and an acquisition phase (see \figurename~\ref{fig::procedure}). 
Each macro phase is composed of three steps, detailed in the following list.
\begin{enumerate}
\item[step 1] fill the tank until the pressure at the gauge reachs a desired value
\item[step 2] wait for the waves produced by the water injection have been damped
\item[step 3] record the stationary value of pressure $P_{max}$ at the gauge
\item[step 4] empty the reservoir of a known amount of water $\Delta V$  for a specified period of time $T_{empty}$ at a constant flow rate equalt to $\Delta V/T_{empty}$
\item[step 5] wait for the waves produced by the water bleeding have been damped
\item[step 6] record the stationary value of pressure $P_{min}$ at the gauge
\end{enumerate}
It should be noted that the values of $P_{max}$, $\Delta V$ and $T_{empty}$ can not be chosen in a completely arbitrary way, but they must be set accordingly with a previsional knowledge of the reservoir, whereas the value of $P_{min}$ is defined by the procedure itself. 
Furthermore, the procedure can be repeated cyclically, \textit{i.e.}, step 1 can be consistently applied after step 6, even if in this case steps 1 to 3 also will be acquisition steps.

It is important to realize that, during step 5, the emptying control is closed and, consequently, the stationary level of the water does not change and the pressure variation acquired at the gauge during this step is due to the damping of the dynamic contribution only. 
This portion of the acquired pressure can be used to model the waves dynamics. For this reason the time period $T_{empty}$ must be set large enough to ensure that the waves have spanned all their dynamic during the step 4.
Afterward, the waves model can be used to clean the output pressure $P_{out}$ acquired during step 4 from the dynamic contribution, allowing us to derive the relationship between the volume $V(P_{qst})$ and the quasi-stationary pressure $P_{qst}$.
Once this relationship is available, the internal shape $C(p)$ can be estimeted accordig with \eqref{eq::cp}.
\subsection{An alternative SoC definition}
Now we are ready to come back to the electrical domain. Replacing pressure with voltage and volume with charge we can express the percentage of the total charge $\Delta Q$ available in the voltage range $[V_{min},V_{max}]$ still stored in the cell at the time $t$ by
\begin{equation}
SoC(V_{qst}(t))=\frac{1}{\Delta Q}\int_{V_{min}}^{V_{qst}(t)}\hspace{-0.5cm}C(\nu)d\nu, \quad V_{qst}(t) \in [V_{min},V_{max}]
\label{eq::newsoc}
\end{equation}
Equation \eqref{eq::newsoc} offers an alternative definition of SoC which overcomes most of the limits of \eqref{eq::oldsoc}. In fact, 
\begin{itemize}
\item it does not require the knowledge of the previous history of the cell (\textit{i.e.} the initial conditions)
\item The amount of charge $\Delta Q$ is well related to $V_{max}$ and $V_{min}$.
\item it does not need an accurate current measurement because it is based on voltage only.
\end{itemize}
Moreover, \eqref{eq::newsoc} is based on the function $C(V_{qst})$ which is a physical property of the cell allowing a cell to cell tailoring of SoC definition.

Unfortunately the successful application of \eqref{eq::newsoc} requires the knowledge of the quasi-statical voltage $V_{qst}$ whereas we are able to measure the output voltage $V_{out}$ only.
For this reason we are forced to develop a cell model well suited to employ a state observer in order to isolate the quasi-statical contribution from the dynamical one. 
\section{Proposed cell Model}
\label{sec::model}
A Thevenin equivalent circuit model (ECM) is said to be realistic if it is able to reproduce within a given error the voltages or the currents measured at the real component when it is driven by any current or voltage waveform. The ECM accuracy depends strongly on the choice of the foundational circuit elements \cite{Chua_2003}. 
In fact, trying to model a nonlinear device through linear components implies the introduction of mathematical artifices, as SoC dependent resistors \cite{Huria_2014, Rahmoun_2012, He_2011}, that do not reflect any physical component.

In this paper, each cell has been modeled as a nonlinear two terminal device. According with the arguments discussed in section \ref{sec::analogy}, the voltage measured at the cell terminals has been modeled as the summation of contributions affecting the cell behavior at different time scales.
In order to include in the model a direct dependence between input current $I_{in}$ and output voltage $V_{out}$, an instantaneous contribution $V_{ist}=\hat{V}_{ist}(I_{in})$ has been added to the quasi-stationary $V_{qst}$ and the dynamic $V_{dyn}$ contributions described in section \ref{sec::analogy}.

According with the previous arguments it can be stated that:
\begin{equation}
V_{out}(t)=V_{qst}(t)+V_{dyn}(t)+\hat{V}_{ist}(I_{in}(t))
\label{eq::Vout}
\end{equation}
Relation \eqref{eq::Vout} can be considered as the output equation of the state form representation of the cell model. 
In order to complete the state form representation of the cell model we need to chose appropriate state variables. According with the mechanical analogy description given in section \ref{sec::analogy}, it seems reasonable to select $V_{qst}$ and $V_{dyn}$ to represent the internal state of the cell.

In order to derive the time evolution of $V_{qst}(t)$, we can consider \eqref{eq::cp} in the electric domain. Solving for the voltage variation and taking the time derivative, we obtain
\begin{equation}
\frac{dV_{qst}(t)}{dt}=\frac{1}{C_{qst}(V_{qst}(t))}\frac{dQ(t)}{dt}=\frac{I_{in}(t)}{C_{qst}(V_{qst}(t))}
\label{eq::Vqstdot}
\end{equation}
It should be noted that $C_{qst}(V_{qst}(t))$ in \eqref{eq::Vqstdot} represents the nonlinear stationary capacity of the cell (\textit{i.e.} its `internal shape').  

\begin{figure}[!htbp]
    \psfrag{Vout}[c][t][1]{$V_{out}(t)$}
    \psfrag{Iin}[c][t][1]{$I_{in}(t)$}
    \psfrag{Vdyn}[c][b][1]{$V_{dyn}(t)$}
    \psfrag{Vist}[c][b][1]{$V_{ist}(t)$}
    \psfrag{Vqst}[c][b][1]{$V_{qst}(t)$}
	\centering
	\includegraphics[scale=0.35]{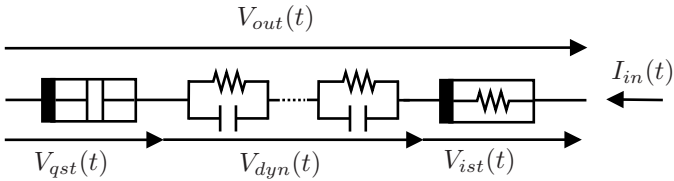}
	\caption{Equivalent circuit of the cell}
	\label{fig::3Blocks}
\end{figure}

In order to keep the model as simple as possible at this stage, the dynamic component $V_{dyn}(t)$ will be considered as the summation of $N$ linear first order filters.
\begin{equation}
V_{dyn}(t)=\sum_{i=1}^{N}V_{i}(t) , \quad \frac{dV_{i}(t)}{dt}=\frac{1}{\tau_{i}}\left(R_{i}I_{in}(t)-V_{i}(t)\right)
\label{eq::Vdyndot}
\end{equation}
The value assumed by the dynamic component after the exhaustion of the transient can be evaluated to be 
\begin{equation}
V_{dyn}^{\infty}=\left(\sum_{i=1}^{N}R_{i}\right) I_{in}(t)
\label{eq::Vdyninf}
\end{equation} 
Equation \eqref{eq::Vout}, \eqref{eq::Vqstdot} and \eqref{eq::Vdyndot} can be interpreted, as shown in \figurename~\ref{fig::3Blocks}, as the series connection of a nonlinear capacitor, a non linear resistor and a cascade of a certain number of RC groups.

\section{Parameter identification}
\label{sec::parameter}
\begin{figure}[!htbp]
    \psfrag{Time}[c][b][0.9]{\scriptsize{Time,~[h]}}
    \psfrag{Volts}[c][t][0.9]{\scriptsize{$V_{out}(t)$,~[V]}}
    \psfrag{Amps}[c][t][0.9]{\scriptsize{$I_{in}(t)$,~[A]}}
    \psfrag{Tempty}[c][b][0.8]{\tiny{$T_{empty}$}}
    \psfrag{step1}[c][b][1]{\tiny{step 1}}
    \psfrag{step2}[c][t][1]{\tiny{step 2}}
    \psfrag{step2b}[c][b][1]{\tiny{step 2}}
    \psfrag{step3}[c][t][1]{\tiny{step 3}}
    \psfrag{step4}[c][t][1]{\tiny{step 4}}
    \psfrag{step5}[c][t][1]{\tiny{step 5}}
    \psfrag{step6}[c][b][1]{\tiny{step 6}}
    \psfrag{DQ}[c][c][1.2]{\tiny{$\Delta Q$}}
    \psfrag{Vmax}[ ][r][0.8]{\tiny{$V_{max}$}}
    \psfrag{Vmin}[ ][l][0.8]{\tiny{$V_{min}$}}
    \psfrag{Vinf}[ ][r][0.8]{\tiny{$V_{dyn}^{\infty}$}}
    \psfrag{yVmax  }[ ][c][0.8]{\tiny{$V_{max}$}}
    \psfrag{yVmin  }[ ][c][0.8]{\tiny{$V_{min}$}}
    \psfrag{Vout(t)}[ ][r][0.7]{\tiny{$V_{out}(t)$}}
    \psfrag{Vdyn(t)+Vmin}[ ][r][0.7]{\tiny{$V_{dyn}(t)+V_{min}$}}
    \psfrag{Vist(t)+Vmin}[ ][r][0.7]{\tiny{$V_{ist}(t)+V_{min}$}}
    \psfrag{Vqst(t)}[ ][r][0.7]{\tiny{$V_{qst}(t)$}}
    \psfrag{Vqstmedio}[r][r][0.8]{\tiny{$\overline{V}_{qst}$}}
	\centering
	\begin{tabular}{c}
		\includegraphics[scale=0.17]{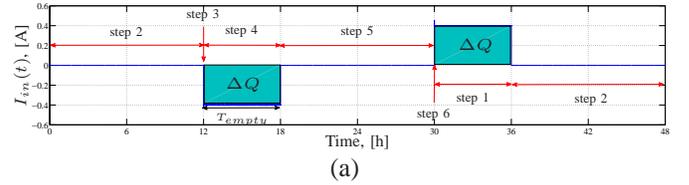} \\
		(a) \\
        \hspace{-0.3cm}	
		\includegraphics[scale=0.173]{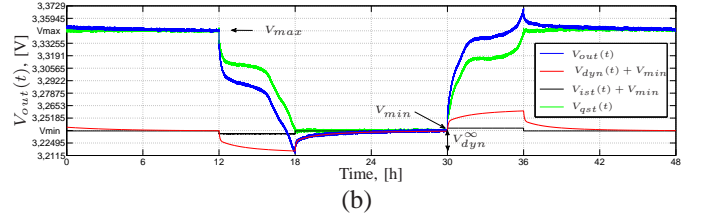} \\
		(b) \\
	\end{tabular}
	\caption{Experimental parameters identification procedure: (a) assigned current profile $I_{in}(t)$, (b) measured voltage profile $V_{out}(t)$ on an AH32113M1Ultra-B Cylindrical cell and its decomposition in the contributions $V_{qst}(t)$, $V_{dyn}(t)$ and $V_{ist}(t)$.}
\label{fig::IandV}
\end{figure}

\begin{figure}[!htbp]
    \psfrag{Amps}[c][t][0.9]{\tiny{$I_{in}(t)$,~[A]}}
    \psfrag{Volts}[c][b][0.9]{\tiny{$V_{out}(t)$,~[V]}}
    \psfrag{Vqst}[c][c][0.9]{\tiny{$V_{max}-V_{min}$}}
    \psfrag{Vdyn}[c][c][0.9]{\tiny{$V_{dyn}^{\infty}$}}
    \psfrag{Vist}[c][c][0.9]{\tiny{$V_{ist}$}}
    \psfrag{Vmin}[c][c][0.9]{\tiny{$V_{min}$}}
    \psfrag{Vmax}[c][c][0.9]{\tiny{$V_{max}$}}
    \centering
		\includegraphics[scale=0.17]{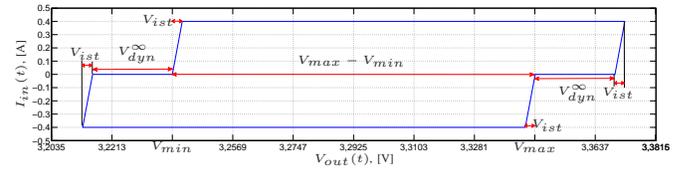}
	\caption{$(V_{out},I_{in})$ parametric plot}
	\label{fig::IinVout}
\end{figure}

In order to determine the model parameters introduced in section \ref{sec::model}, we should apply the characterization procedure described in section \ref{sec::analogy} to determine the reservoir's shape to a specific cell.
The cell taken into consideration in this paper is the A123 Nanophosphate AHR32113M1Ultra-B. The main characteristics of this cell are given in \tablename~\ref{tab::A123_32113_specs}.
%
%
%
\begin{table}[htbph]
\centering
\begin{tabular}{cc}
Variable & Setting  \\
\hline
\hline
Cell Dimensions (mm) & 32x113 \\
Cell Weight (g) & 205 \\
Cell Capacity (nominal/minimum, Ah) & ~4.4 \\
Energy Content (nominal, Wh) & 14.6 \\
Discharge Power (nominal, W) & 550 \\
Voltage (nominal, V) & 3.3\\
Specific Power (nominal, W/kg) & 2700 \\
Specific Energy (nominal, Wh/kg) & 71 \\
Energy Density (nominal, Wh/L) & 161 \\
Operating Temperature(°C) & -30 to 55 \\
Storage Temperature (°C) & -40 to 60 \\
\end{tabular}
\caption{A123 Nanophosphate AHR32113M1Ultra-B specifications.}
\label{tab::A123_32113_specs}
\end{table}
The experimental parameters identification is based on the procedure described in the mechanical domain in section \ref{sec::analogy}. 
The current input profile and the output voltage measured on the cell taken into consideration are shown in \figurename~\ref{fig::IandV} part (a) and (b), respectively.
The values of circuital components related to the instantaneous portion of the voltage can be obtained by interpreting the current $I_{in}(t)$ and the voltage $V_{out}(t)$ as coordinates and the time $t$ as a parameter, so that an interesting plot can be obtained in the $(I,V)$ plane, as shown in \figurename~\ref{fig::IinVout}. 
It is easy to understand that the oblique lines in \figurename~\ref{fig::IinVout} represent a simultaneous jump in both the current and the voltage that correspond to the instantaneous contribution of the output voltage.
The slope of those lines represents the value of the internal resistance at the specified current values. Similar tests performed at different current values can be used to derive the characteristic $\hat{V}_{ist}(I_{in})$ representing, from  circuital point of view, a nonlinear resistor.
On the midline of \figurename~\ref{fig::IinVout}, together with the instantaneous contributions, the variation ranges of ${V}_{qst}$ and ${V}_{dyn}$ are labeled with $V_{max}-V_{min}$ and $V_{dyn}^{\infty}$, respectively.

The values of circuital components related to the dynamic portion of the output voltage can be obtained by fitting the measured voltage shown in \figurename~\ref{fig::IandV} part (b) acquired in correspondence of zero current after the current pulse has been removed (\textit{i.e.} from hour 18 to 30 and 36 to 48). As expected, these portions of the $V_{out}(t)$ curve correspond to a low-pass behavior and it can be identified with a certain number of $RC$ groups connected in series as stated in \eqref{eq::Vdyndot}. Moreover, as stated by \eqref{eq::Vdyninf}, the total resistive contribution can be estimated from the difference $V_{dyn}^{\infty}$ between the output voltage measured at the 30-th and at the 18-th hour. 
According with \eqref{eq::Vout}, once both the instantaneous and the dynamic components $V_{ist}$ and $V_{dyn}$ of the circuits have been identified the quasi-static portion $V_{qst}$ of output voltage $V_{out}$ can be obtained by subtraction. 
\begin{equation}
V_{qst}(t)=V_{out}(t)-V_{dyn}(t)-\hat{V}_{ist}(I_{in}(t))
\label{eq::Vqst}
\end{equation}
The proposed technique is quite different from the identification techniques present in the litterature based on linear combinations of nonlinear functions \cite{Plett_2004_b,Paschero_2010_b}, point by point \cite{Birkl_2015} or on averaging charge and discharge curves \cite{Hu_2011, Baronti_2015}.

The three components of $V_{out}(t)$ have been plotted together on \figurename~\ref{fig::IinVout} part (b). In order to improve the readability of the plot both $V_{ist}(t)$ and $V_{dyn}(t)$ have been added to $V_{min}$.
From \figurename~\ref{fig::IinVout}, it is possible to realize that the only voltage contribution for zero current is the dynamical one \textit{i.e.}  $V_{out}(t)=V_{dyn}(t)$ for $I_{in}(t)=0$. This property is really useful to identify $V_{dyn}$ (\textit{i.e.} the wave dynamics). Moreover, it should be noted that when subject to positive and to negative current pulses, $V_{dyn}$ exhibits a non symmetrical behavior resulting in an error in the open circuit voltage estimation when the mean value between the charge and the discharge branches is considered \cite{Hu_2011, Baronti_2015}.

\begin{figure}[!htbp]
    \psfrag{Q}[c][t][0.9]{\scriptsize{$Q(V_{qst}) / \Delta Q$}}
    \psfrag{Volts}[c][b][0.9]{\scriptsize{$V_{qst}$,~[V]}}
    \psfrag{Vmax}[c][b][0.9]{\scriptsize{$V_{max}$}}
    \psfrag{Vmin}[c][b][0.9]{\scriptsize{$V_{min}$}}
    \psfrag{Vn}[c][b][0.9]{\scriptsize{$V_{n}$}}
	\centering
		\includegraphics[scale=0.17]{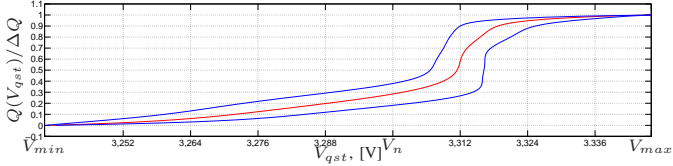}
	\caption{$Q(V_{qst})$ characteristic. Mean value in red.}
	\label{fig::OCVSOC}
\end{figure}

\begin{figure}[!htbp]
    \psfrag{Vqst}[c][b][0.9]{\scriptsize{$V_{qst}$,~[V]}}
    \psfrag{Vmin}[c][b][0.9]{\scriptsize{$V_{min}$}}
    \psfrag{Vmax}[c][b][0.9]{\scriptsize{$V_{max}$}}
    \psfrag{Vn}[c][b][0.9]{\scriptsize{$V_{n}$}}
    \psfrag{DVn}[c][c][0.9]{\scriptsize{$\Delta V_{n}$}}
    \psfrag{C}[c][t][0.9]{\scriptsize{$C_{qst}(V_{qst})$,~[F]}}
	\centering
		\includegraphics[scale=0.17]{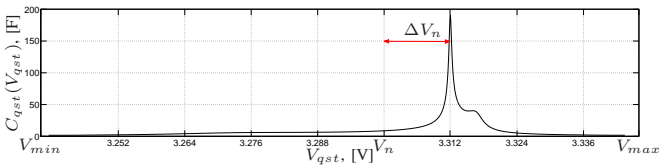}
	\caption{Estimated nonlinear statical capacity}
	\label{fig::capqst}
\end{figure}

According with the literature, the estimated $V_{qst}(t)$ exhibits different behaviour during charge and discharge. In fact, plotting the current integral $Q(t)$ versus this portion of the output voltage, the characteristic $Q(V_{qst})$ shown in \figurename~\ref{fig::OCVSOC} is obtained. 
According with the arguments given in section \ref{sec::analogy}, for applying \eqref{eq::newsoc} we need to estimate the statical capacitance of the cell. The hysteretic behavior of the cell produces different capacitances (shapes) for the charge and the discharge process. In order to simplify the model, we are forced to consider an average between the charge and the discharge behaviors. Consequently, the static capacitance of the cell $C_{qst}(V_{qst})$ can be estimated applying \eqref{eq::cp} to the mean curve shown in red in \figurename~\ref{fig::OCVSOC}. The curve obtained for  $C_{qst}(V_{qst})$ is plotted in \figurename~\ref{fig::capqst}.
From the plot it is evident that most of the charge is accumulated around a single statical voltage value corresponding to the maximum value of the capacitance. This argument can be used as a constructive definition of the nominal voltage $V_n$ and it strengthens the choice made in this paper to use $V_{min}$ and $V_{max}$ values closer to $V_{n}$ with respect to the values given by the constructor. Moreover, it is interesting to note that following the proposed procedure there is just a small shift $\Delta V_{n}$ between the nominal voltage given by the cell constructor and the peak of the estimated statical capacitance. 
\section{Multi-Cell Extended Kalman Filter}
\label{sec::multi_cell_ekf}

\subsection{Framework}
\label{sec::multi_cell_ekf_framework}

\begin{figure}[!htbp]
	\centering
		\includegraphics[scale=0.375]{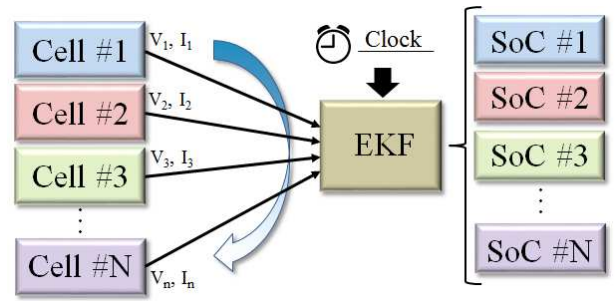}
	\caption{General block diagram of the Multi-Cell EKF framework.}
	\label{fig::multi_cell_framework_diagram}
\end{figure}

As well known in  the literature, the EKF is a powerful method to estimate the states of a nonlinear system such as internal parameters and SoC of a battery. 
However, this approach presents several weak points:
(i) the precision of the model implemented into the observer can heavily influence the estimation precision; (ii) depending on the application the time convergence to reach a reasonable estimation can be too long; (iii) the initialization of the auto-covariance Q matrix can be difficult; (iv) the EKF is able to estimate the states of only one system (\textit{i.e.} a single cell). 
In order to improve the robustness and the reliability of this methodology, regarding points (i), (ii) and (iii), it is possible to increase the complexity of the battery model, perform a better initial condition estimation and empirically find the best Q matrix values, respectively.
Considering point (iv), to our best knowledge, to date has not been proposed any approach allowing the estimation of the states of multiple cells using only one observer (\textit{i.e.} EKF).
Thus, in a real system composed by dozens of cells, as many EKFs have to be implemented, exacerbating the complexity of the system and the used memory space.
For these reasons, in the last year the University of Rome Sapienza and the Ohio State University have worked together in order to find a way to develop a novel methodology defined Multi-Cell EKF.
The idea is quite simple and it is based on the well known round-robin strategy, where time slices are assigned to each process in equal portions and in circular order, handling all processes without priority.
Starting from this idea, it has been implemented a framework that allows a single EKF to perform the states estimation of different batteries.
Indeed, considering the ``slow'' evolution of voltage and current during normal applications, it is reasonable to expect similarly slow changes of battery conditions.
The block diagram explaining the round-robin approach is reported in \figurename~\ref{fig::multi_cell_framework_diagram}.
Even though the round-robin strategy is well known, some relevant comment can be made.
Using this novel approach, the estimation of cells either with the same chemistry or not residing either in the same battery pack or in different ones is possible.
Indeed, as shown in \figurename~\ref{fig::multi_cell_framework_diagram}, the EKF receives only voltage and current values ignoring any other information of the cells (\textit{i.e.} chemistry and position in the battery pack).
In order to simplify the description of this novel Multi-Cell EKF, we consider a battery pack constituted by the connection in series of cells all having the same chemistry.

\begin{figure}[!htbp]
	\centering
		\includegraphics[scale=0.285]{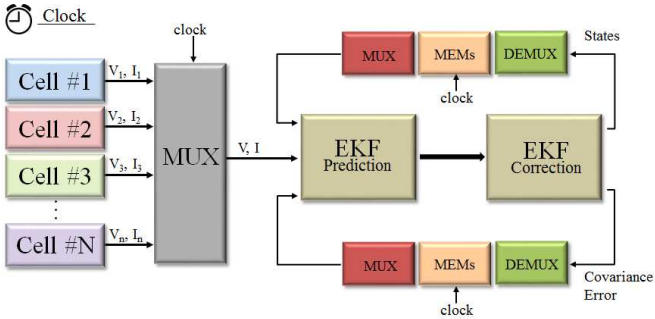}
	\caption{Block diagram of the developed Multi-Cell EKF framework.}
	\label{fig::multi_cell_framework_diagram_2}
\end{figure}

Even though the round-robin concept shown in \figurename~\ref{fig::multi_cell_framework_diagram} can appear quite simple, its correct implementation is not trivial due to the recursive structure of the observer. 
In fact the EKF algorithm is composed by two steps defined \textit{Prediction} and \textit{Correction}.
The first step involves projecting both the most recent states estimation and an estimate of the error covariance (from the previous time period) forwards in time to compute a predicted (or a-priori) estimate of the states at the current time. The second step involves correcting the predicted states calculated in the first step by incorporating the most recent process measurement to generate an updated (or a-posteriori) state estimation.
Thus, applying the round-robin strategy, for each time slice the 
\textit{Prediction} step has to be fed by suitable data from the  \textit{Correction} step at the previous time period. 
Along this line, the general framework block diagram shown in \figurename~\ref{fig::multi_cell_framework_diagram} has been modified as shown in \figurename~\ref{fig::multi_cell_framework_diagram_2}, where the actual EKF has been used as an engine to perform calculations, whereas the framework has been used to change the cell under test every time slice.

\begin{figure}[!htbp]
	\centering
		\includegraphics[scale=0.275]{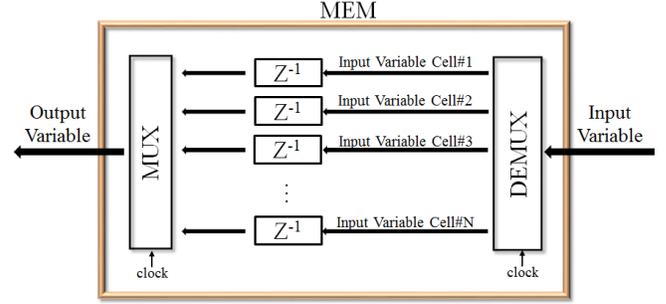}
	\caption{Inner part of the MEM block.}
	\label{fig::mem_diagram}
\end{figure}

In the block diagram reported in \figurename~\ref{fig::multi_cell_framework_diagram_2}, it is possible to see how the round-robin structure is implemented using multiplexers (MUXs), demultiplexers (DEMUXs), memories (MEMs) driven by a common clock signal. 
MUXs have been used to select suitable data for each time slice, MEMs have been used to store real time data of each cell, DEMUXs have been used to properly separate variables contained in arrays and the clock has been used to set the sampling time $T_{slice}$.
The EKF Prediction and Correction blocks are two common processing blocks with known latencies. This modular design allows to modify, tune and improve the latter blocks, without modifying the remainder of the system.  
Furthermore, as mentioned before in cases where cells have different chemistry, it is possible to customize, for instance, the battery internal parameters increasing the versatility of the proposed method.
The MUXs and DEMUXs blocks shown in \figurename~\ref{fig::multi_cell_framework_diagram_2} perform standard functions, so they can be implemented using built-in blocks or custom functions according to the programming environment available. 
These blocks have been used to separate or concatenate the variables to or from the MEMs blocks that represent the states and the auto-covariance Q matrix, respectively.
MEMs blocks, used to properly store data, have been implemented using a custom solution.
These blocks, playing a vital role, constitute the most important part of the entire Multi-Cell EKF structure.
A single MEM block has to be used for every EKF state (\textit{i.e.} $V_{qst}$ and $V_{dyn}$).
The block diagram of a single MEM is reported in \figurename~\ref{fig::mem_diagram}.
For each time step, the single variable coming from the \textit{Correction} block is properly addressed to the specific delay block and stored, whereas its previous value is sent as output variable to the \textit{Prediction} block.
This updating operation is performed on a single cell at the time, driven by the clock rate, while all the other data are left unchanged.
\subsection{Determining the maximum number of cells}
\label{sec::maximum_number}  
\begin{figure}[!htbp]
	\centering
	\psfrag{Mod}[c][t][0.9]{\scriptsize{$|I_{in}(\omega)|$,~[As]}}
	\psfrag{freq}[c][b][0.9]{\scriptsize{$\omega$,[Hz]}}
	\psfrag{Amps}[c][t][0.9]{\scriptsize{$I_{in}(t)$,~[A]}}
    \psfrag{Time}[c][b][0.9]{\scriptsize{$t$,[s]}}
	\begin{tabular}{c}
	   \includegraphics[scale=0.23]{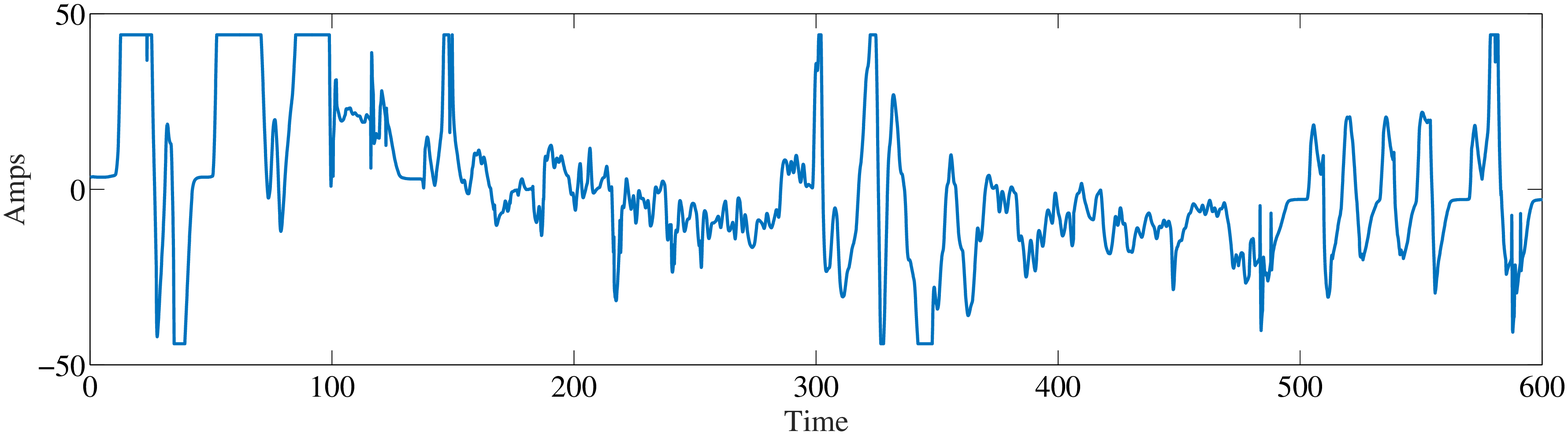}\\
	   (a)\\
	   \includegraphics[scale=0.23]{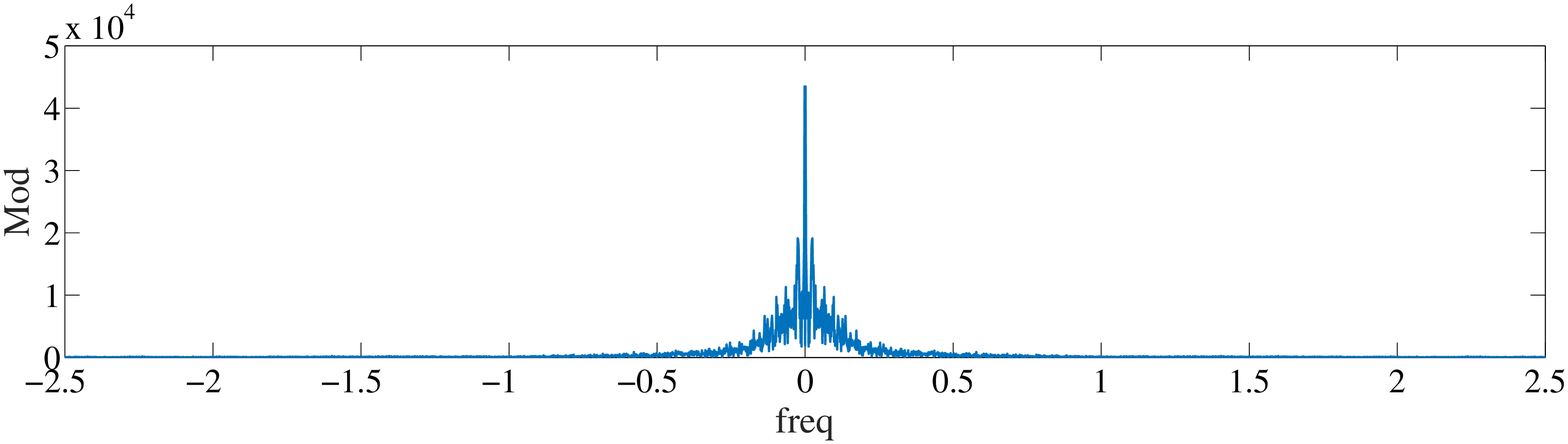}\\
	   (b)\\
	\end{tabular}
	\caption{Current profile derived from the standard US06 driving cycle: (a) time evolution, (b) spectrum.}
	\label{fig::US06_current}
\end{figure}

As discussed in the previous section, the proposed framework gives the possibility to perform states estimation of different cells implementing a single EKF. 
Aimed to determine the maximum number of cells that the Multi-Cell EKF can deal with in a real scenario, a realistic current profile has been used.
The current profile, shown in \figurename~\ref{fig::US06_current} part~(a), is derived from the well known US06 driving profile, where a maximum current limitation of 44~[A] (10 C-Rate) is considered. This constraint has been applied only for safety reasons, based on the cell characteristics of the cell used in the actual test (see \tablename~\ref{tab::A123_32113_specs}).
The spectrum of the input signal $I_{in}(t)$ is reported in \figurename~\ref{fig::US06_current} part~(b), where the maximum frequency can be cautiously fixed to 2~Hz.
The maximum number of the cells that the Multi-Cell EKF can manage is dictated by the input signal maximum frequency (\textit{i.e.} current and voltage) as well described in the Nyquist-Shannon sampling theorem.
By setting the time slot $T_{slot}$ (slice time of each process in the round-robin strategy) equal to 10~ms to allow the correct computation of the data with the available hardware, the maximum number of the cells supported by the Multi-Cell EKF is computed as follows:
\begin{equation}
\#Cells T_{slot} \leq \frac{1}{2 f_{max}} \Rightarrow \#Cells  \leq \frac{1}{2 f_{max}T_{slot}}= 25
\label{eq::number_cells}
\end{equation}
According with \eqref{eq::number_cells}, where a current profile derived from a standard US06 driving cycle is applied, the Multi-Cell EKF is able to perform the SoC estimation up to 25 cells. 
Considering that hundreds of cells are normally present in a regular battery pack, the reduction of the computing hardware costs using this approach is remarkable.
\section{Tests and Results}
\label{sec::results} 
\subsection{Model Validation}
\begin{figure}[htbph]
\centering
    \psfrag{Volts}[c][t][0.9]{\scriptsize{$V_{out}$,~[V]}}
    \psfrag{Volts1}[c][t][0.9]{\scriptsize{Error,~[mV]}}
    \psfrag{Time}[c][b][0.9]{\scriptsize{Time,[h]}}
\begin{tabular}{c}
\includegraphics[scale=.24, angle=0]{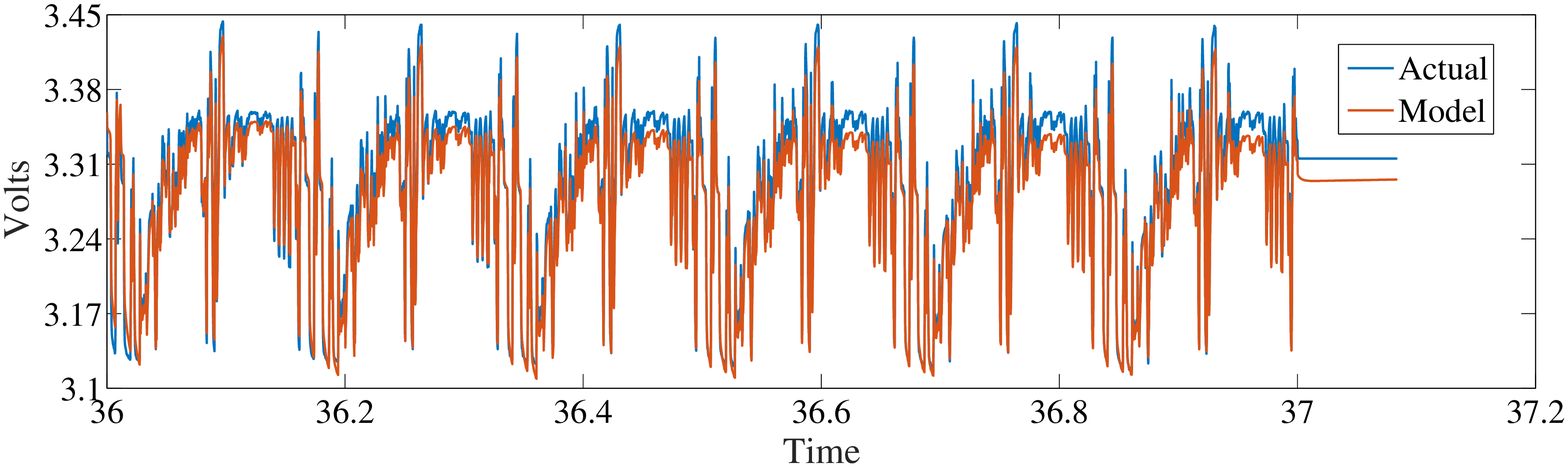} \\ 
(a)\\
\includegraphics[scale=.24, angle=0]{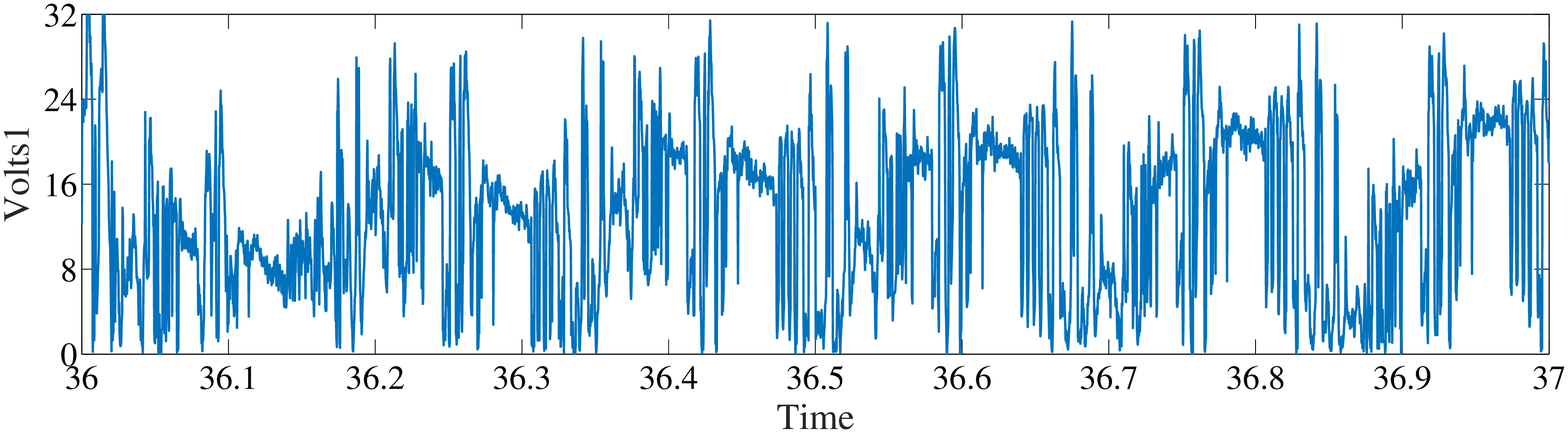} \\
(b)\\
\end{tabular}
\caption{Comparison between model and actual voltage: (a) Comparison, (b) absolute difference.}
\label{fig::voltage_comparison}    
\end{figure}

In order to validate the cell model described in sections \ref{sec::model} and \ref{sec::parameter}, we use a custom current profile having a total time length of about 37 hours. It is composed by:
\begin{itemize}
\item 6 hours of discharging at a constant 0.4~[A] rate
\item 12 hours of rest
\item 6 hours of charging at a constant 0.4~[A] rate
\item 12 hours of rest
\item 1 hour of variable current composed by the concatenation of 6 US06 current profiles shown in \figurename~\ref{fig::US06_current}~part~(a)
\end{itemize}
During the first 36 hours the current profile is quite similar to the one shown in \figurename~\ref{fig::IandV}~part~(a) used in the identification procedure. We take this choice to verify that the model works well in a slow dynamic situation. Conversely, for the latter hour of the current profile, we take a high dynamic behavior in order to stress the model.
A comparison between the actual voltage acquired from the cell and the one estimated by the model and their absolute error is reported in \figurename~\ref{fig::voltage_comparison} for the latter hour.
Inspecting \figurename~\ref{fig::voltage_comparison}~part~(b) a few considerations can be made. 
First of all it should be noted that the voltage produced by the model is almost always less than the actual one. This behavior is probably due to the inability of the model in describing hysteresis phenomena. Even if the results achieved with the actual model are encouraging, we believe that a better modeling of the hysteresis phenomena should reduce the mean value of the error. 
\subsection{Multi-Cell EKF Validation}
\begin{figure}[htbph]
\centering
\psfrag{Q}[c][t][0.9]{\scriptsize{SoC}}
\psfrag{Time}[c][b][0.9]{\scriptsize{Time,[h]}}
\begin{tabular}{c}
\includegraphics[scale=.24, angle=0]{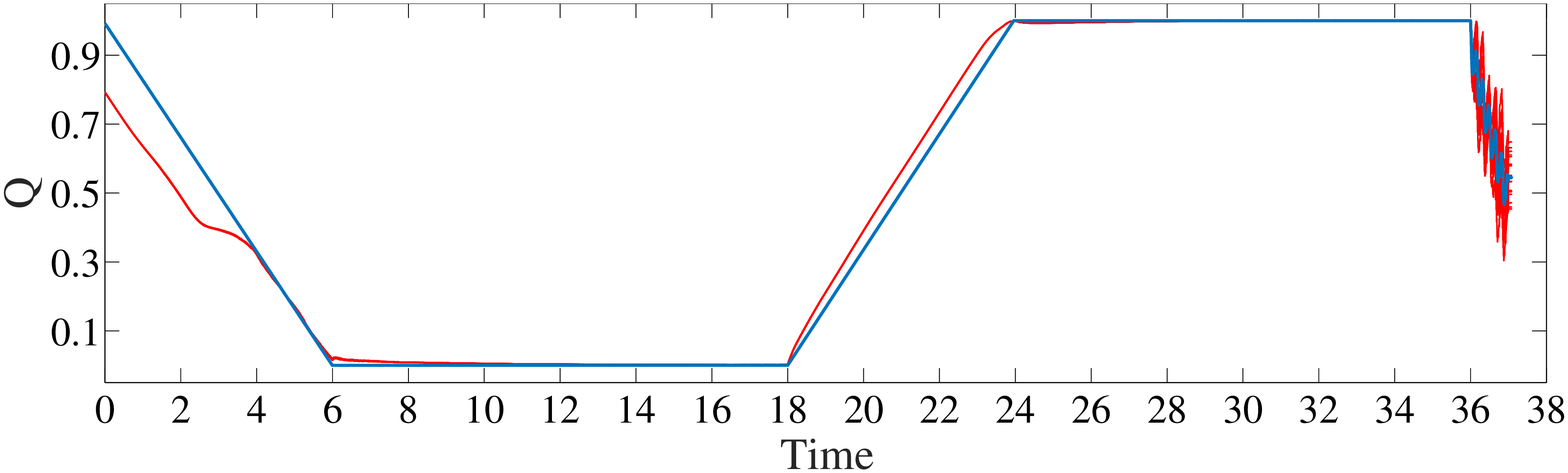} \\ 
(a)\\
\includegraphics[scale=.25, angle=0]{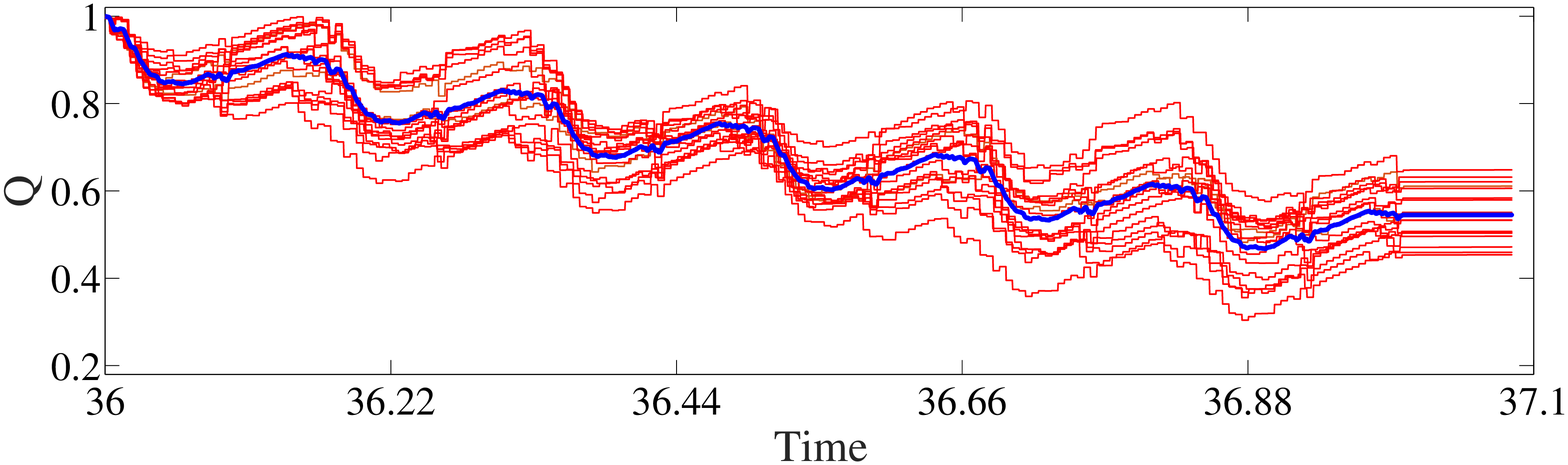} \\
(b)\\
\end{tabular}
\caption{Comparison between Coulomb counting (blue) and the SoC of the 18 cells estimated by the Multi-Cell EKF (red): (a) full test, (b) zoom of the US06 section.}
\label{fig::soc_initial_100_parallel}    
\end{figure}

Once the model is validated, in order to verify the effectiveness of the Multi-Cell EKF in a real operative scenario, we decided to exploit the same current profile used to validate the model, on a prototype battery pack composed by 18 A123 Nanophosphate AHR32113M1Ultra-B connected in series. Before starting the test all the 18 cells have been prepared to ensure the condition $V_{qst}=V_{max}$.
The Multi-Cell EKF registers have been set with an initial $V_{qst}$ value corresponding to about the $80\%$ of the total storable charge. Being the actual cells charged at $V_{qst}=V_{max}$ the initial error on the state estimation is about $20\%$.
The results of the SoC estimation performed by the Multi-Cell EKF are reported in \figurename~\ref{fig::soc_initial_100_parallel}, where the Coulomb counting has been used as SoC reference.
As visible in the \figurename~\ref{fig::soc_initial_100_parallel}~part~(a), the Multi-Cell EKF is able to engage the SoC reference as a traditional EKF showing how the proposed framework does not alter its functioning. In the part (b) of the same figure is depicted a zoom of the latter hour of the test showing that in high dynamical condition the SoC estimated of each cell is slightly different from the others with a maximum difference up to $10\%$ with the reference.
This behavior is expected since in an operative scenario the cells differ one to each other in terms of internal parameters, age, etc..
In order to perform a better estimation, authors are actually working on an improved round-robin strategy able to include cell to cell tailored parameters. 
\section{Conclusions}
\label{sec::conclusion}
An alternative SoC definition, based on a mechanical analogy, has been introduced in order to avercome a few limits of the standard definition. Based on the same analogy, a novel equivalent nonlinear circuit model of a cell has been derived together with a suitable parameters identification procedure which avoid the use of non physical components such as SoC dependant resistors. A multi-cell SoC estimation using a single EKF making use of a round-robin approach has been described and a design procedure to extimate the maximum number of cell it is possible to menage with a singol EKF filter has been derived. The tests conducted to validate the model show a good accuracy in SoC estimation, even in very challenging conditions. Similar test conducted on a prototype battery pack composed of 18 cells connected in series show the ability of the proposed EKF to prerform an accurate multi-cell SoC estimation. 
\bibliographystyle{IEEEtran}
\bibliography{../../../reference_database/references_db}
\begin{IEEEbiography}[{\includegraphics[width=1in,height=1.25in,clip,keepaspectratio]{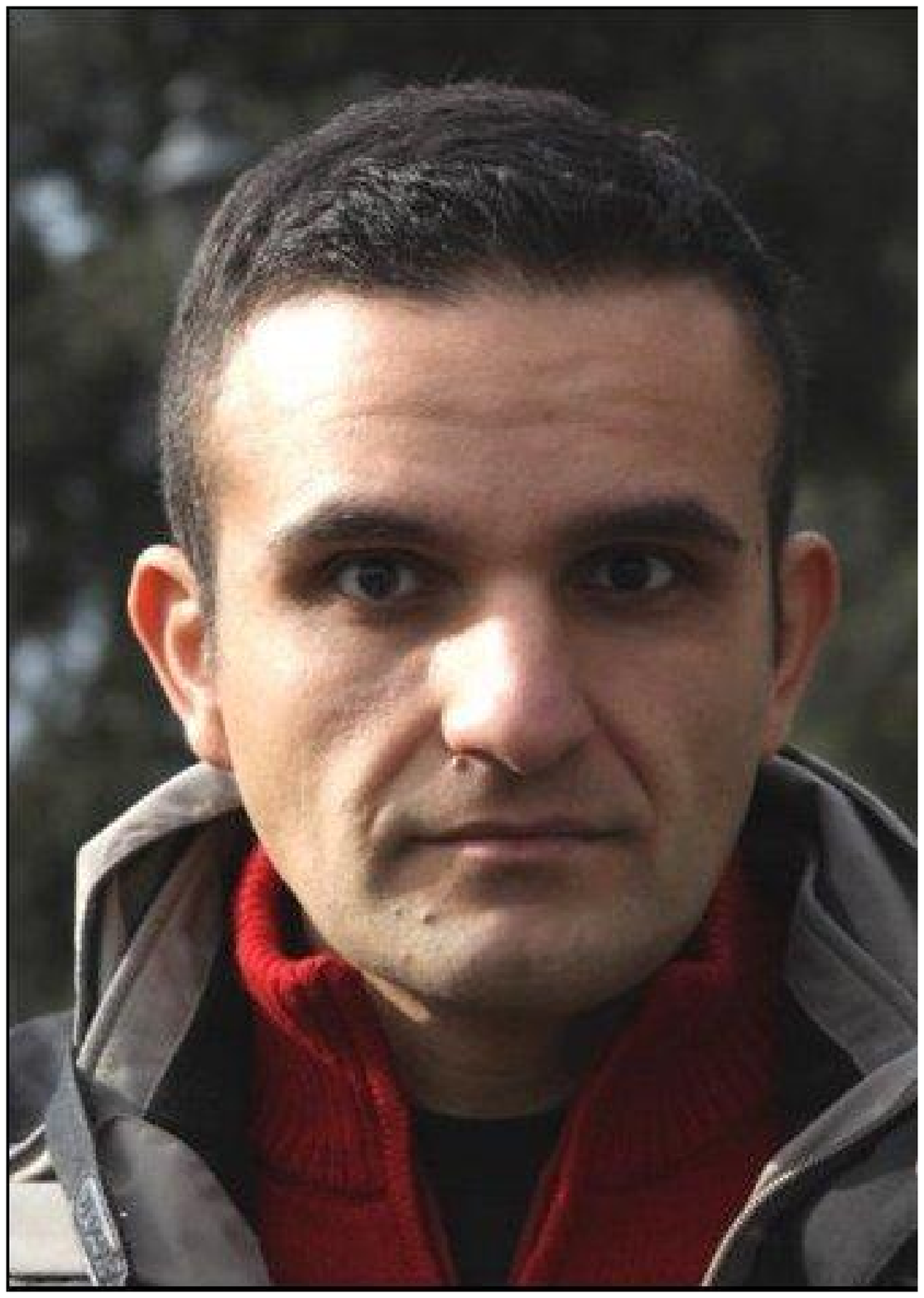}}]{Maurizio Paschero}  is a post doctoral research associate at the Information Engineering, Electronics and Telecommunications Department of the University of Rome "La Sapienza" since September 2008, where he works in the Polo per la Mobilit\`{a} Sostenibile (POMOS) Laboratories.
He received his M.S in Electronic Engineering 2003 and the Ph.D in Information and Communication Engineering in 2006 from the University "La Sapienza" of Rome and the Ph.D in Mechanical Engineering in 2008 from Virginia Polytechnic Insitute and State University.
His major fields of interest include Smart Grids, circuital modeling of multi-physic systems, intelligent signal processing, and battery modeling.
\end{IEEEbiography}
\begin{IEEEbiography}[{\includegraphics[width=1in,height=1.25in,clip,keepaspectratio]{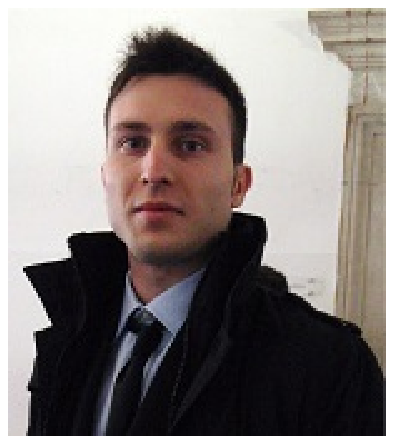}}]{Gian Luca Storti} is a post doctoral research associate at Center for Automotive Research of Ohio State University since May 2015.
He received his B.S and his M.S in Electronic Engineering from "La Sapienza" University, Rome, Italy, in 2008 and 2011, respectively.
He received his Ph.D. in the Information Engineering, Electronics 
and Telecommunications Department in 2015 from the same university.
His major field of interest include the Smart Grids, intelligent signal processing, modeling and control 
of hybrid powertrain, design of automotive electrical system and battery modeling.
\end{IEEEbiography}
\begin{IEEEbiography}[{\includegraphics[width=1in,height=1.25in,clip,keepaspectratio]{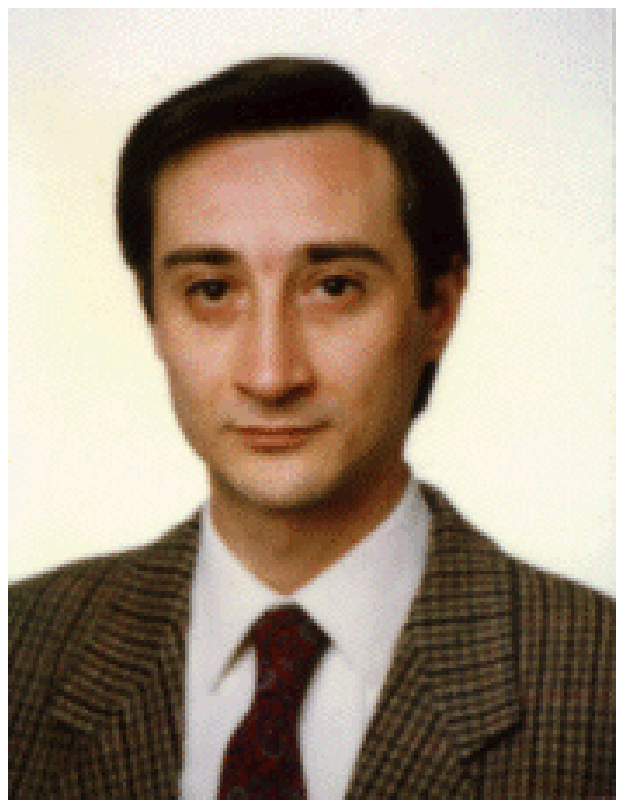}}]{Antonello Rizzi} 
(S’98-M’04) received the Ph.D. in Information and Communication Engineering in 2000, from the University of Rome ”La Sapienza”. In September 2000 he joined the DIET Department of the University "La Sapienza" of Rome as an Assistant Professor. His major research interests are in the area of non-linear circuits and systems, Computational Intelligence and Pattern Recognition, including advanced Granular Computing techniques for Data Mining and Knowledge Discovery. Since 2008, he serves as the scientiﬁc coordinator and technical director of the R\&D activities in the Intelligent Systems Laboratory within the `Polo per la Mobilit\`{a} Sostenibile' (POMOS) Laboratories, where he is working on the design of intelligent systems for sustainable mobility, smart grids and micro-grids modeling and control, and battery management systems. Dr. Rizzi (co-)authored more than 110 international journal/conference articles and book chapters.
\end{IEEEbiography}
\begin{IEEEbiography}[{\includegraphics[width=1in,height=1.25in,clip,keepaspectratio]{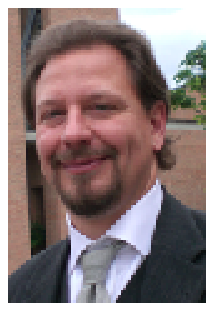}}]{Fabio Massimo Frattale Mascioli} 
Prof. Fabio Massimo Frattale Mascioli received his MS and PhD in Information and Communication Engineering in 1989 and 1995, from the University "La Sapienza" of Rome. In 1996, he joined the DIET Department of the University "La Sapienza" of Rome as Assistant Professor. He was promoted to Associate Professor of Circuit Theory in 2000 and to Full Professor in 2011. His research interest mainly regards neural networks and neuro-fuzzy systems and their applications to clustering, classification and function approximation problems, circuit modeling for vibration damping, energy conversion systems and electric and hybrid vehicles. He is author or co-author of more than 150 papers. Since 2007, he serves as scientific director of the `Polo per la Mobilit\`{a} Sostenibile' (POMOS) Laboratories, DIET Department.
\end{IEEEbiography}
\begin{IEEEbiography}[{\includegraphics[width=1in,height=1.25in,clip,keepaspectratio]{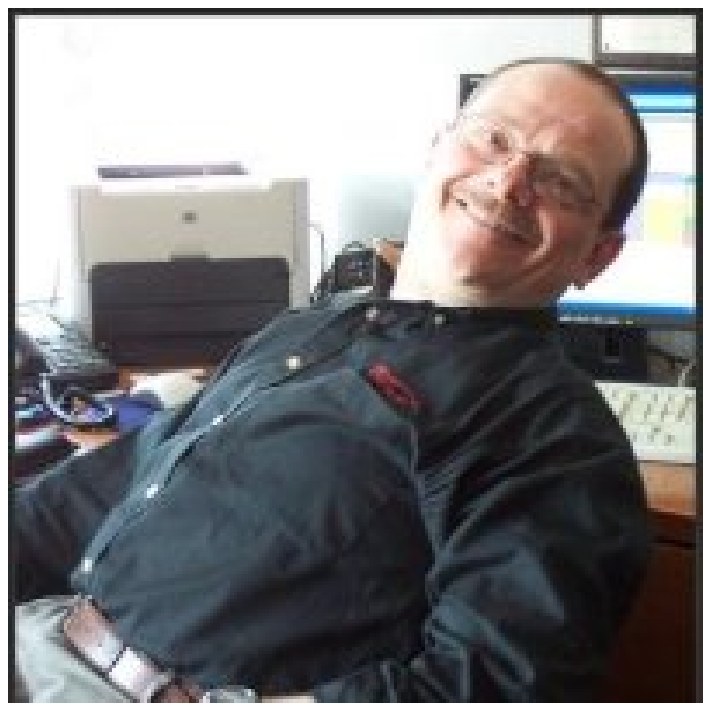}}]{Giorgio Rizzoni} 
received his BS, MS and PhD in Electrical and Computer Engineering in 1980, 1982 and 1986, from the University of Michigan.  
Between 1986 and 1990 he was a post-doctoral fellow, and then a lecturer and assistant research scientist at the University of Michigan. In 1990 he joined the Department of Mechanical (now Mechanical and Aerospace) Engineering at Ohio State as an Assistant Professor. He was promoted to Associate Professor in 1995 and to Professor in 2000.  In 1999 he was appointed director of the Center for Automotive Research.  Since 2002 he has held the Ford Motor Company Chair in Electromechanical Systems, and has also been appointed Professor in the Department of Electrical and Computer Engineering. Prof. Rizzoni's specialization is in dynamic systems and control, and his research activities are related to sustainable and safe mobility. He is a Fellow of the Institute of Electrical and Electronic Engineers and Fellow of the Society of Automotive Engineers, and has received numerous teaching and research awards, including the Stanley Harrison Award for Excellence in Engineering Education and the NSF Presidential Young Investigator Award.
\end{IEEEbiography}

\end{document}